\documentstyle[preprint,prb,aps]{revtex}
\begin{document}
\draft
\title{
Adsorbed 3d transition metal  atoms and  clusters  
on  Au(111): Signatures derived from one-electron calculations.  }

\author{Mariana Weissmann $^{a}$ and Ana Maria Llois $^{a,b}$}
\address{$^a$ Departamento de Fisica, Comision Nacional de Energia Atomica,
Avda. del Libertador 8250, 1429 Buenos Aires, Argentina\\
$^b$ Departamento de Fisica, FCEyN-UBA, Ciudad Universitaria, 1428 
Buenos Aires, Argentina} 

\date{\today}
\maketitle
\begin{abstract}
The spectroscopic characteristics of systems with adsorbed $d$ impurities
on noble metal surfaces should depend on the number and geometric
arrangement of the adsorbed atoms and also on their $d$ band filling.
Recent experiments using scanning tunneling microscopy 
have probed the electronic structure of all $3d$
transition metal impurities and also of Co dimers adsorbed on Au(111),
 providing a rich variety
of results. In this contribution we correlate those experimental results
with  ab-initio calculations and try to establish necessary conditions for
observing a Kondo resonance when  using the single impurity
Anderson model.
We find that the relevant orbitals at the STM tip position, when it
is on top of an impurity, are the $d$ orbitals with $m=0$  and that
the energy of these levels with respect to the Fermi energy determines
the possibility of observing a spectroscopic feature due to the impurity.

\end{abstract}

The electronic structure of $3d$ transition-metal impurities on the (111)  
surface of copper and gold  has been studied recently 
with scanning tunneling microscopy \cite{Crommie1,Crommie2,Crommie3,elipse}.
The first measurements, performed only for Co impurities,
gave rise to a direct spectroscopic observation of the 
Kondo resonance for an isolated impurity \cite{Crommie1}. Further experiments 
with Co dimers \cite{Crommie2} and with
other $3d$ transition metal atoms \cite{Crommie3} show that there is a different
behavior of $dI/dV$ for different impurity $d$ filling, thus varying significantly
across the $3d$ series in the periodic table. These results provide  a rich
phenomenology that we  believe 
can be correlated with results of LSDA calculations. 

In a  previous paper \cite{nuestro} we calculated  the electronic 
densities of states of a periodic system that simulates the  experimental
situation described above. It 
consisted of repeated slabs of five layers of Au(111) with 
Co atoms deposited on both sides of them 
forming a dilute adsorbed layer. For this purpose we
used the FP-LAPW method and the LSDA approximation 
with the WIEN97 code \cite{Blaha}. The main result of that paper was that the
minority spin contribution to the density of states due to the $3d$ orbitals
of Co with  $m=0$ 
  is a very narrow band located
 precisely at $E_F$, which could be related to the observed experimental feature.

In the present work we have extended the calculation
to other transition metals and also to dimers. The
unit cell  is the same for all the impurity
cases, it has five layers of Au with three atoms in each layer. One
impurity atom is located on each side of the slab at one of the hollow
sites of the Au(111) slab structure forming a $\sqrt{3}\times\sqrt{3}$ 
adsorbed layer.
The slabs are  separated by enough empty space so as to
simulate non interacting surfaces.  The distance between impurity
atoms on the same plane is 5 \AA, and their distance to the surfaces is
relaxed so as to minimize the energy.

Magnetism can be quantified in these  calculations 
by giving the
 magnetic moment inside the
muffin tin spheres or  the $d$ band splitting for the different impurities.
 We show these results in Table I,  they agree with previous
calculations for similar adsorbed systems \cite{otros}.
The given splittings are those corresponding to  the separation between 
majority and minority centers of
 the $m=0$
narrow band, that we assume is the more relevant one. Both magnitudes,
$\mu$ and $E_{up}-E_{dn}$, are
seen to be roughly proportional.
In order to simplify the analysis and to compare with
results from model Hamiltonian calculations
we have made the working assumption that it is enough to consider 
just the  single $3d$ level with $m=0$ in order to
 correlate the calculated results with the STM experimental ones.
These experiments were performed in a small energy range  around the
Fermy energy of Au, using a bias voltage of less than 0.1 eV. Our
calculations indicate that only
 the majority band of V and the minority
band of Co fall within that range. One would then expect similar experimental
results for both impurities,
 but this is not the case \cite{Crommie3}.

Local densities of states at the impurity sites, of $d$ symmetry and 
for  $m=0$ in particular, are shown in Figs. 1 to 6. Also shown
in that figure are the
charge densities in the vacuum region, plotted 
along the plane
perpendicular to the slab at the impurity site,  corresponding to two small
energy windows close to $E_F$. According to the usual theory of STM
spectroscopy \cite{Tersoff} the tunneling current is directly related
to the charge density at the tip position. We see from Figs. 1 to 6 that the
magnitude of this
charge density along the normal to the surface is in all cases 
significantly larger
for the energy window  that contains the $3d$ orbital with $m=0$ 
symmetry, namely $w1$. 
The largest hybridization with the conduction electron sea  therefore
takes place 
at the energies where the $m=0$ symmetry provides the largest contribution.
Thus,  tunneling conductance should be  essentially
affected by the   presence of these symmetries on the impurity atoms 
if they happen to fall in the narrow energy window where the experiments
are made. 
 This effect should be observed both for minority
or majority spins, if the energy conditions were given. 

When analyzing each case in more detail we see that for Ti
the majority spin electron density in window $w1$  is
about one order of magnitude larger than the minority contribution at the
tip site (4 to 5 \AA,  from the surface) and also larger than the contributions
of both majority and minority character in the other window ($w2$).
 This agrees very well with the observed increase 
in the STM signal
when the bias voltage sweeps energies from below to above the Fermi level,
going across this last one.
We believe this reflects the increased density of states just above the
Fermi level, when electrons start to flow from the tip into the empty states added by
the impurity. Because of the experimental width of this feature  it is possibly
 not a manifestation of the Kondo effect.
The second largest value for the ratio of charge density in window $w1$ 
versus the one in window $w2$ is
obtained for Ni, and may be interpreted in a similar way as that of Ti,
although the  magnitudes are much smaller. This system could not show a
Kondo effect as it is non-magnetic.
For Cr and Fe the relevant states are not close to $E_f$
so that their influence due to 
hybridization should 
not be visible. 
The cases of Co and V, where we find a narrow band peaked at $E_f$,  present
very similar values of the ratio of charge densities in both windows,being  
this ratio of the order of 2. Besides this, the absolute charge density values 
are also very small.  We shall analyze below in which cases
one could expect a signature of the Kondo peak.
 
An important one-electron feature, not considered in our calculations, is
the interaction of the adsorbed atoms with the (111) surface state at the
$\Gamma$ point of 
noble metals.
This state is  responsible for the appearance of standing
wave patterns in quantum corrals, and  also possibly of 
recently observed mirages
 \cite{elipse}. Our representation of the
surface by slabs that are only five layers thick would give a poor
approximation of the surface state  but the periodicity of the adsorbed
atoms, that are not too far away from each other, probably destroys
 the surface state completely.
 As a first approximation we  assume that this interaction will not 
depend on 
 $d$ band filling. 

To study the importance of many body effects not considered in 
the LSDA calculations  
the single impurity Anderson Hamiltonian (SIAM) is one of the usual approaches
\cite{SIAM}.
 It is known to be equivalent, for most sets of parameters, to the Kondo model
Hamiltonian, through the Schrieffer-Wolf transformation \cite{Kondo}.
It consists of a conduction band, that provides the Fermi energy or
zero energy, and a single magnetic impurity level at $E_0$. This level
interacts with the band, thus acquiring a width $\delta$, and electrons
with opposite spins occupying that level feel a Coulomb repulsion $U$.
The mean field solution of this model Hamiltonian results in two narrow bands,
centered at $E_0$ and $E_0+U$, which could be parametrized for the different
adsorbed systems using the
results of our LSDA calculations, for example setting for $E_0$ and $E_0+U$ 
 $E_{up}$
and $E_{dn}$ of Table I, respectively.
The many body solution of this model Hamiltonian  
gives rise to a narrow peak at the Fermi level of the
spectral function.
This Kondo peak is always present at T=0, except when $E_0$
 (or $E_0+U$) is closer
than $\delta$ to the Fermi energy. In this situation, the mixed valence case,
the many body feature is masked inside the broader one-electron peak
 \cite{Hewson}.
The possibility of actually seing the  Kondo peak  experimentally is
limited by the exponential dependence of the Kondo temperature on the
separation of $E_0$ (or $E_0+U$) from the Fermi energy.
This may be the reason why it is not seen for most of the $3d$ transition
metal impurities. For Co and V the present LSDA calculations 
give the $m=0$ level very close to $E_F$ and would therefore predict the 
possibility of observing  a Kondo peak in both cases.
However, in the case of Co  the density of states for $m=0$ shows narrower features
that for V, this last one spreading more uniformly within an energy range
of more than 0.2 eV. This  could explain why no special STM feature
is observed in V.
 It is well known that 
the LDA approximation enhances the width of 
hybridization peaks, so that we are not able to decide if the observed feature
in the case of Co
is essentially many body (Kondo)  or not. The temperature dependence would of
course help to answer this question.
In any case, only when one of the two levels, at $E_0$ or at $E_0+U$, is close to
$E_F$ there is a possibility of observing this narrow feature.

When two Co impurities are brought close together \cite{Crommie2} the
experimental feature of the single Co impurity dissappears. In a simple
tight-binding model one would expect a splitting of both the majority
and the minority spin levels  due to the interaction between the two impurities,
 giving rise to 
 pairs of bonding
and antibonding states. To perform an LSDA calculation similar to the previous
one 
requires a much larger unit cell in order to prevent each dimer from interacting 
with the others. For this last reason we performed a simplified calculation,
 using only one 
monolayer of Au(111)
and depositing dimer rows on one side of it. The two atoms in the dimer were
placed at a separation of
$2.87 \AA$, in two next nearest neighbour hollow sites of the surface, 
and atoms in different dimers were situated at least $5 \AA$  apart.
For Co dimers in the ferromagnetic configuration we found a dimer magnetic 
moment of $4 \mu_B$ and the
$m=0$ majority levels appear at -3.10 and -1.27 eV, while
the minority levels are at -1.54 and +0.43 eV with respect to the Fermi
level of gold. 
None of the levels is within the experimental energy range and for 
this reason 
the Kondo temperature would be an order of magnitude or more lower than for an
isolated Co impurity. 
 The antiferromagnetic
configuration of the dimer has a very small energy difference with the
ferromagnetic one, 
and  the $3d$  levels are also far from $E_F$.
  Therefore, the reason for
not observing the narrow feature at $E_f$ in the dimer case may be either that
the Kondo temperature is too low or that the system is antiferromagnetic,
in which case there would be no Kondo peak.

In summary, we have related results of LSDA calculations with features
 appearing near the
Fermi level when transition metal atoms or clusters are adsorbed on Au(111).
The proximity of the impurity $d$ level with $m=0$  to the Fermi energy of 
the conduction band
seems to be a necessary but not sufficient condition for observing an
abrupt change
in the tunneling current.

\acknowledgements
We thank Dr.Javier Guevara for helpful discussions, and
 acknowledge CONICET and SECYT (grant PICT 03-00105-02043) of Argentina
for partial support.

\begin{table}
\begin{tabular}{ccccccc} 
             &     Ti   &    V    &    Cr   &  Fe    &    Co   &  Ni\\ 

$\mu$        &   0.83   &   2.45  &   3.43  &  3.03  &  1.80   &  0.00\\
$E_{up}-E_f$ &   0.25   &  -0.02  &  -0.56  & -2.50  & -1.65   & -0.30\\		 
$E_{dn}-E_f$ &   0.88   &   1.75  &   2.36  &  0.30  & -0.06   & -0.30\\ 
\end{tabular}
\caption{
Magnetic moments (in Bohr magnetons) inside the muffin-tin sphere  and 
position of the narrow energy bands with $m=0$ with respect to the
Fermi energy (in  eV).  
}
\label{Table I}
\end{table}
\begin {figure}
\caption{
 Ti atom adsorbed on the Au(111) surface.
Left) Partial densities of states in the muffin-tin sphere of the impurity. The full lines
show all states of $d$ symmetry while the dashed lines show the $m=0$ orbital
only.
Energies are in eV and referred to the Fermi level of Au.
Right) Decay of the charge density into the vacuum, along the perpendicular
to the (111) surface, starting on top of the adsorbed atom. Only 
states in two energy windows, close to the Fermi energy,
are reported. Window $w1$ is chosen so that it contains the $d$ states
with $m=0$, window $w2$ is next to it and both are 0.2 eV wide.
The full line indicates majority spin density contribution within
$w1$, the dashed line  minority spin density contribution within $w1$,
the dash-dot line minority contribution within $w2$ and
the dotted line  majority contribution within $w2$}.
\label{Fig. 1}
\end{figure}
\begin {figure}
\caption{
Same as Fig. 1  but for a V impurity.}
\label{Fig. 2}
\end{figure}
\begin {figure}
\caption{
Same as Fig. 1  but for a Cr impurity.}
\label{Fig. 3}
\end{figure}
\begin {figure}
\caption{
Same as Fig. 1  but for a Fe impurity.}
\label{Fig. 4}
\end{figure}
\begin {figure}
\caption{
Same as Fig. 1  but for a Co impurity.}
\label{Fig. 5}
\end{figure}
\begin {figure}
\caption{
Same as Fig. 1  but for a Ni impurity.}
\label{Fig. 6}
\end{figure}
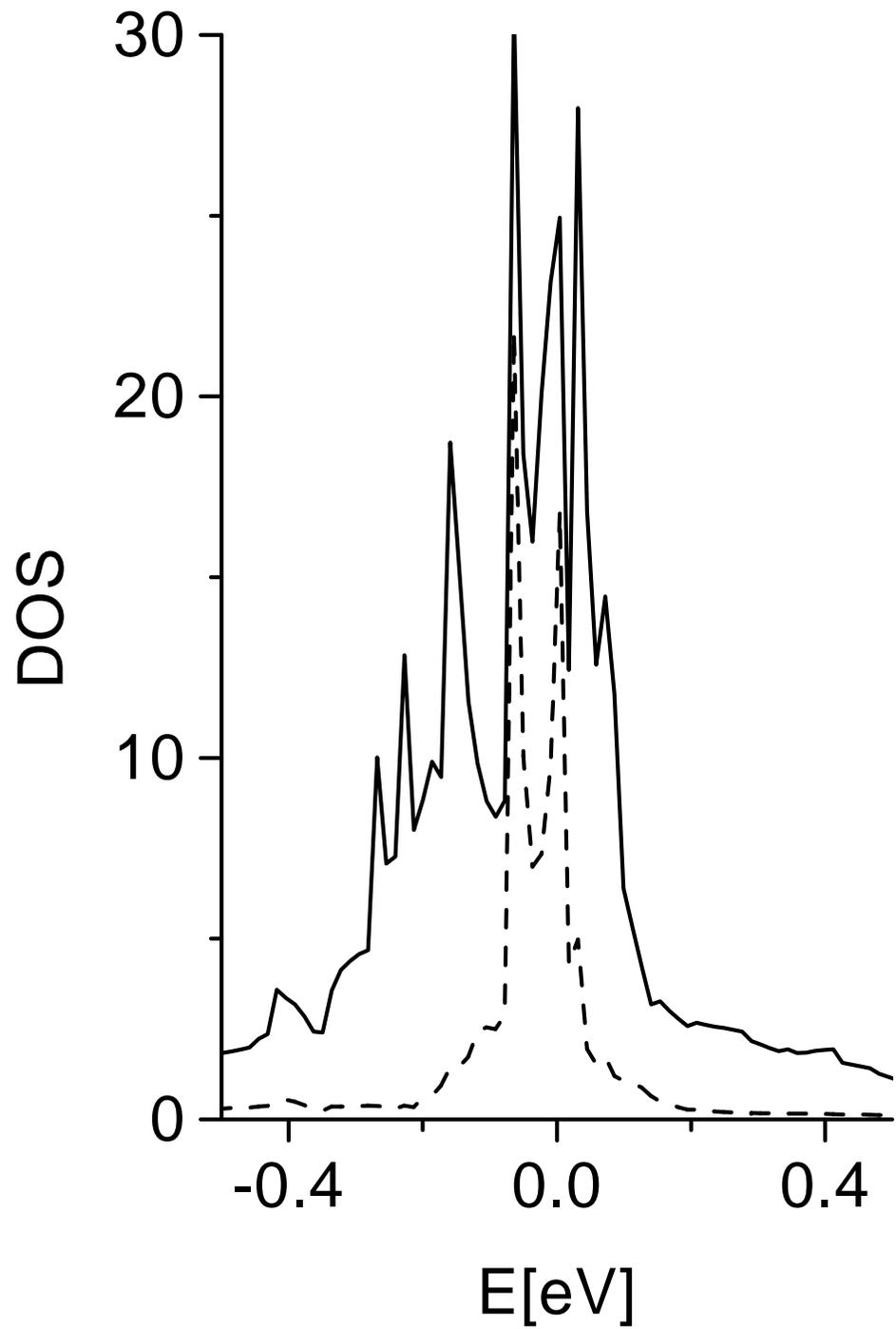
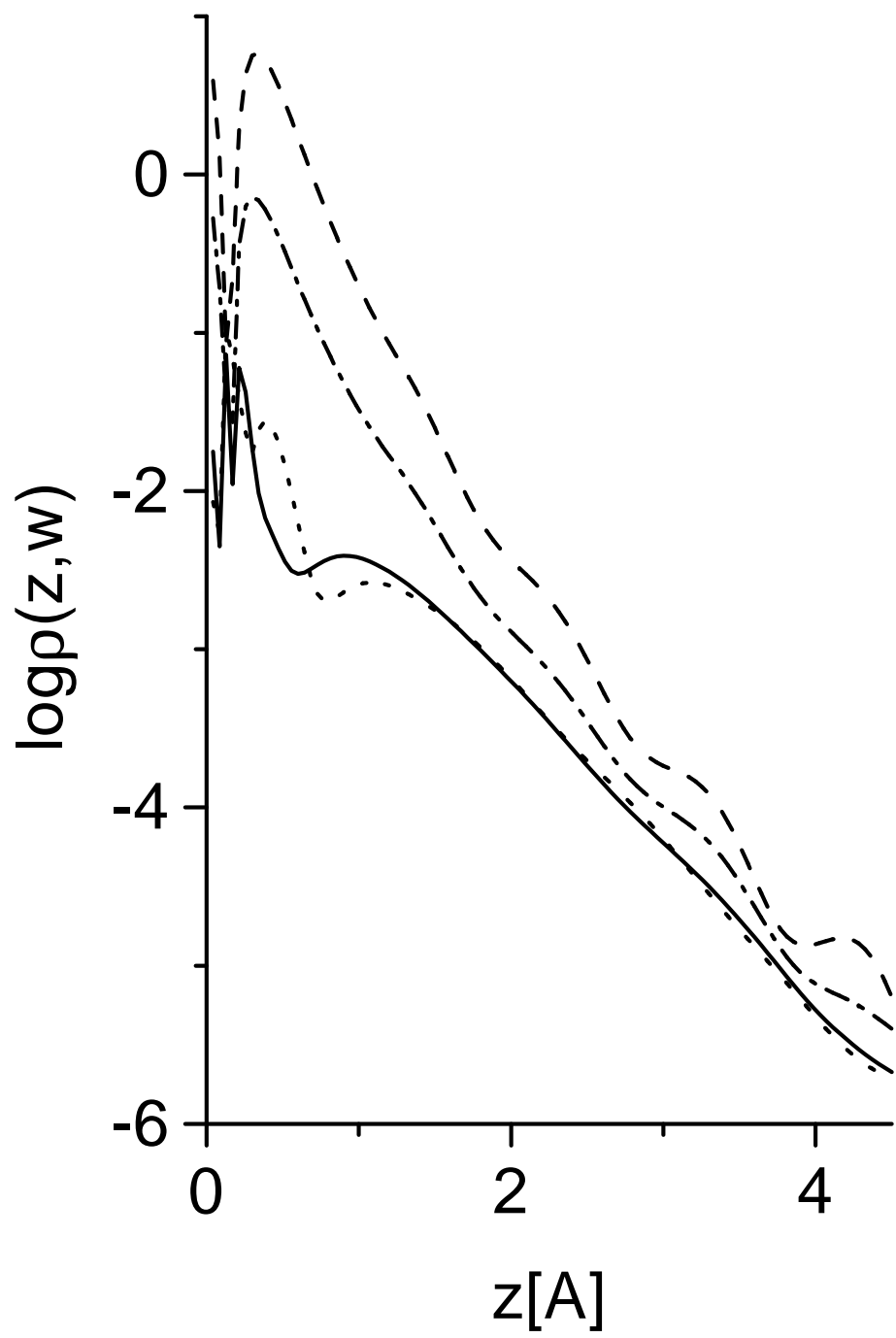
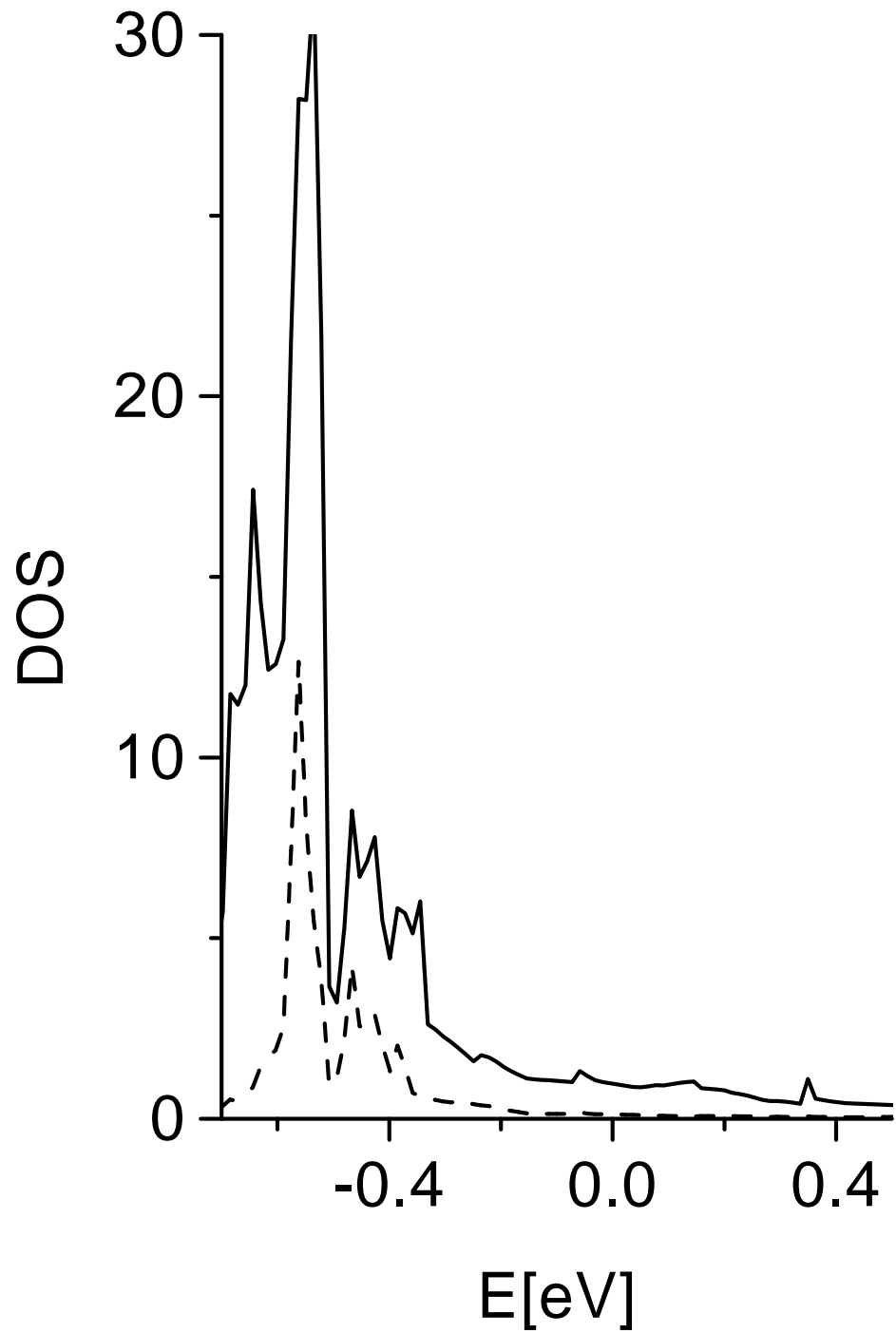
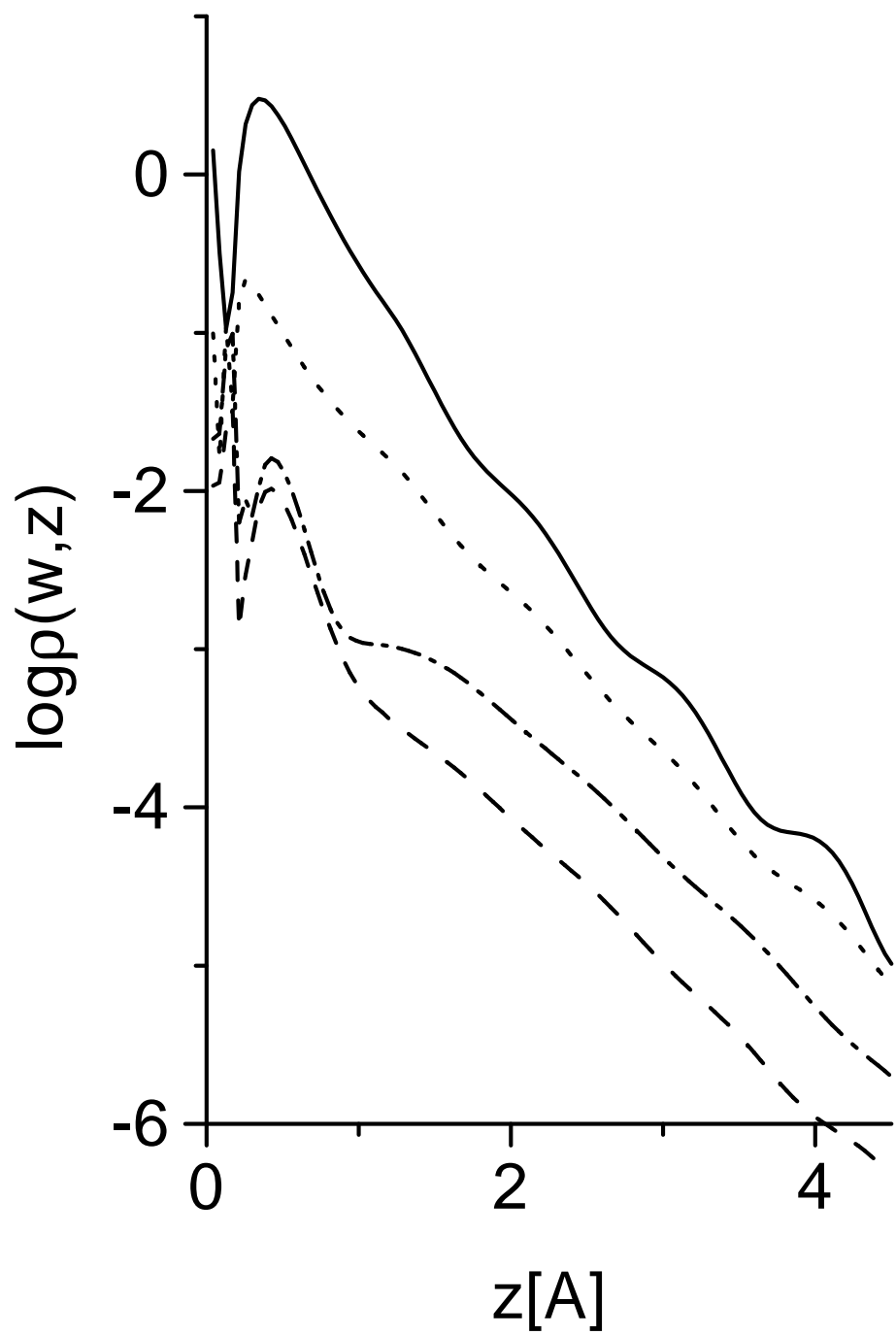
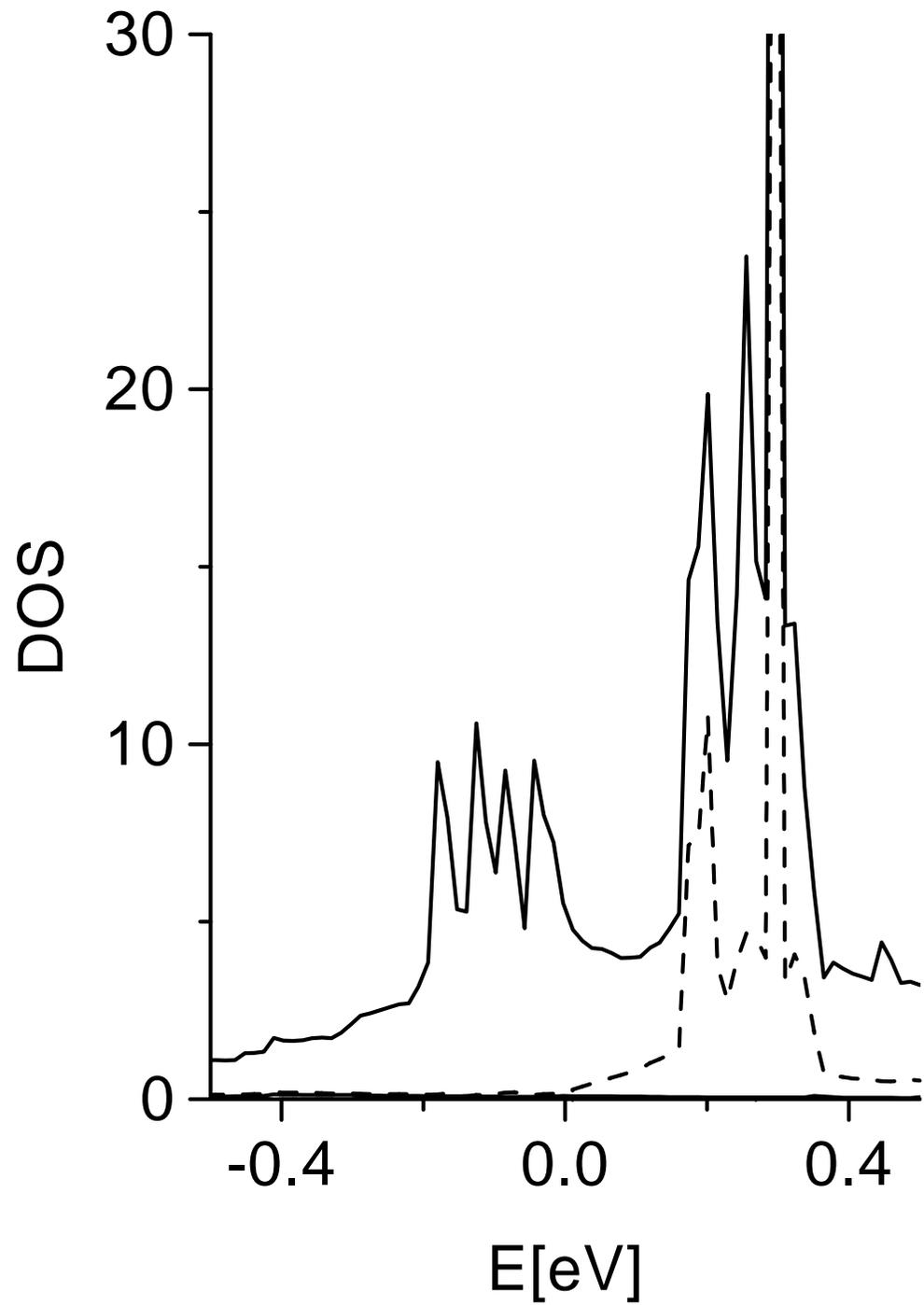
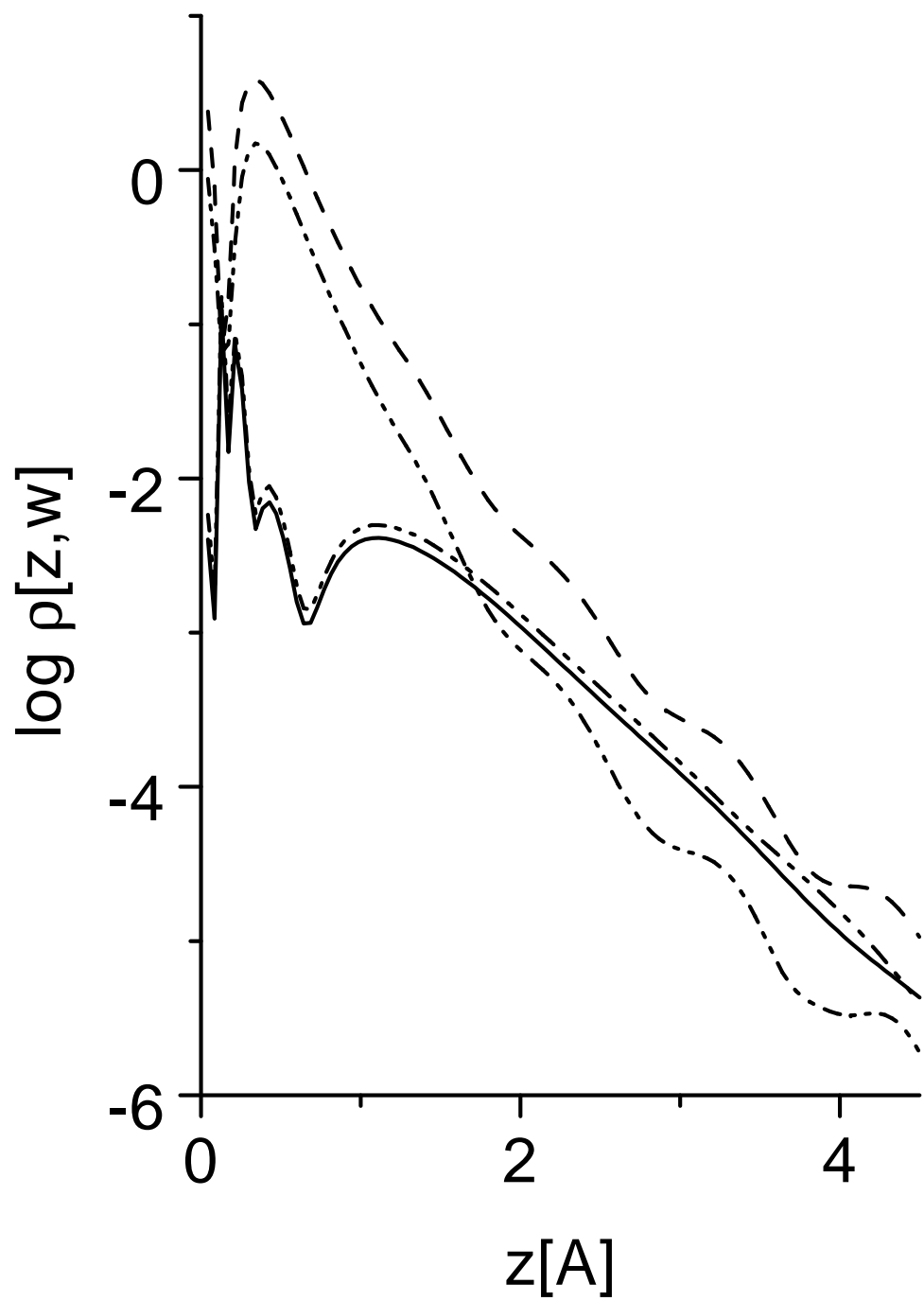
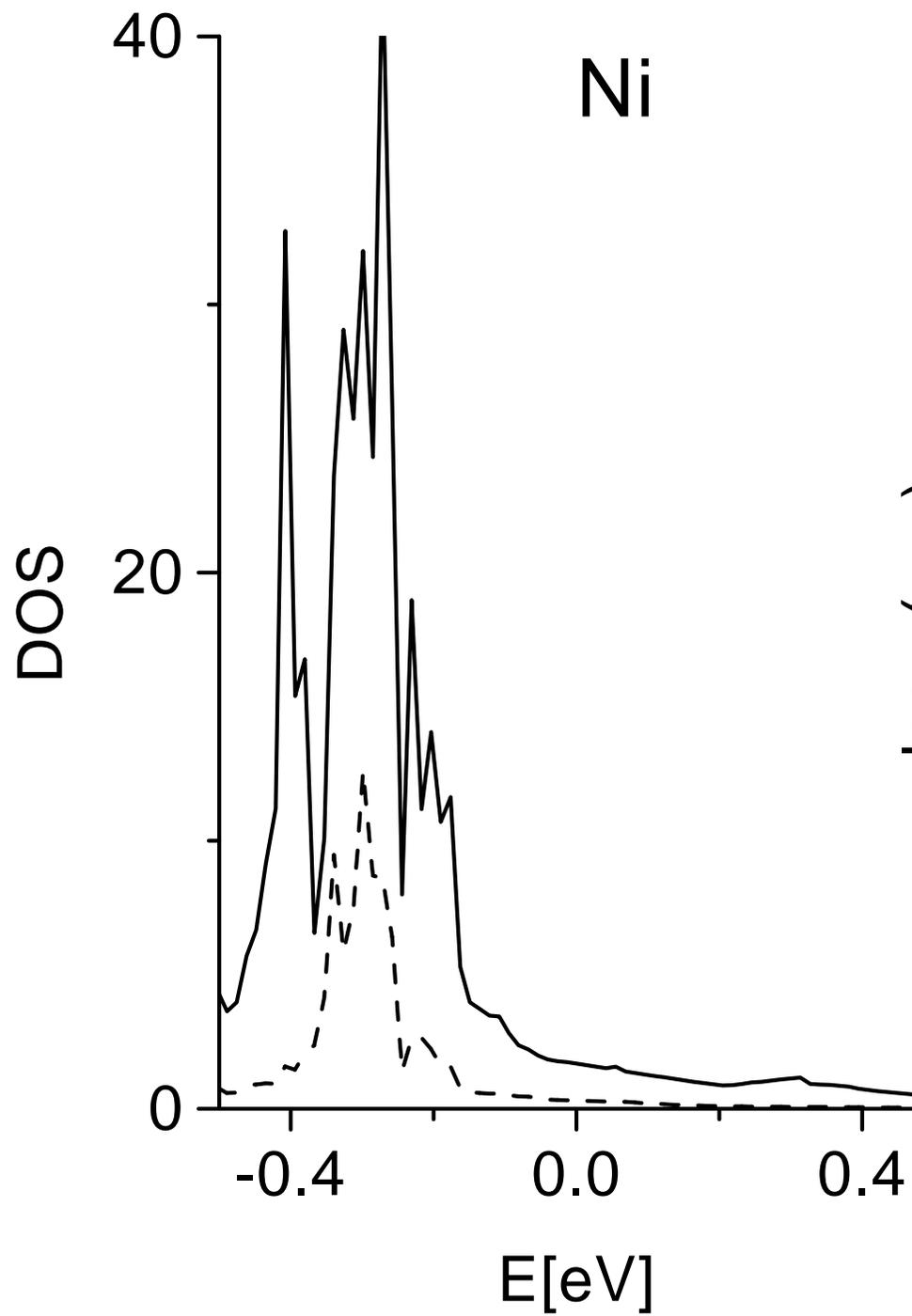
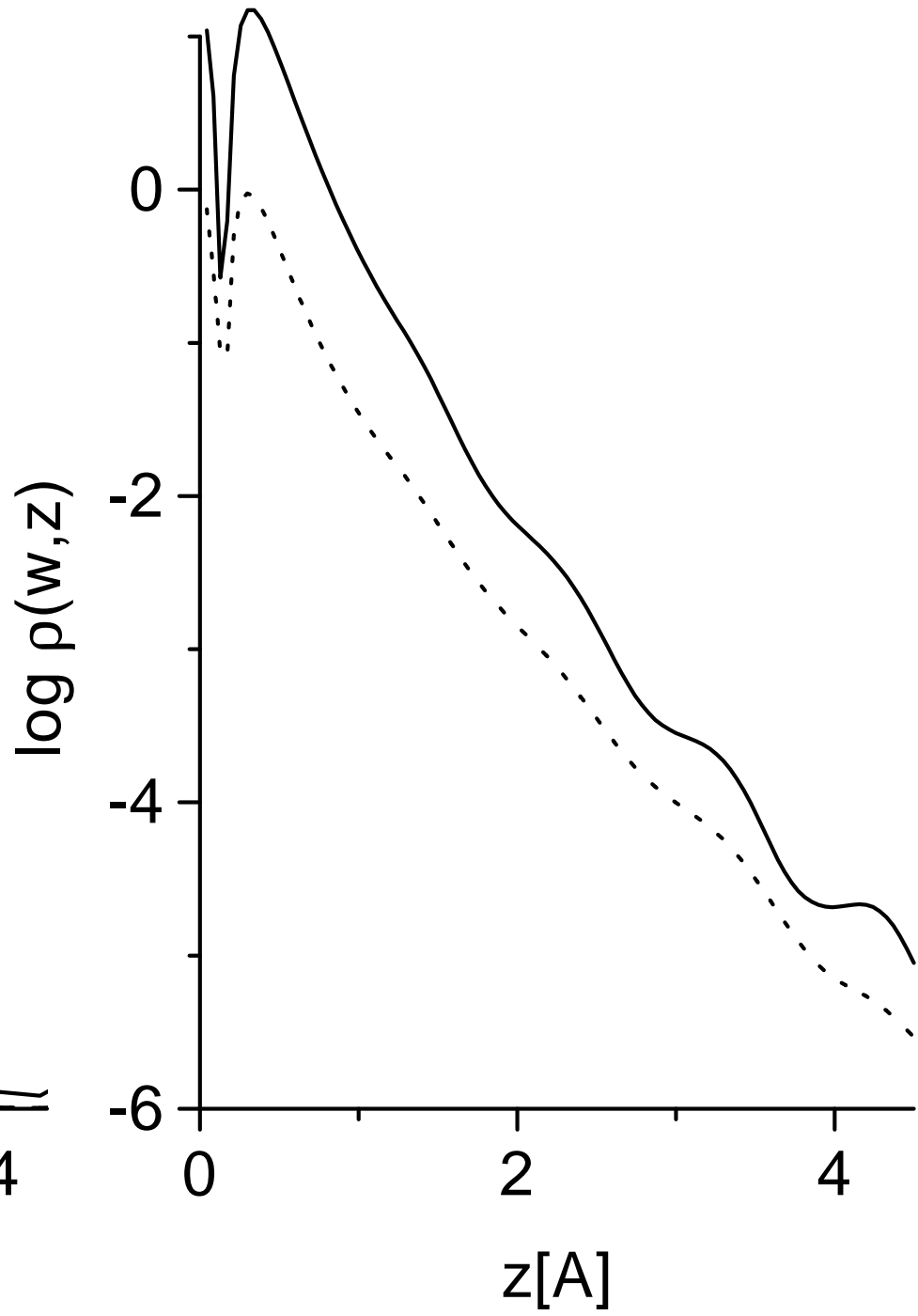
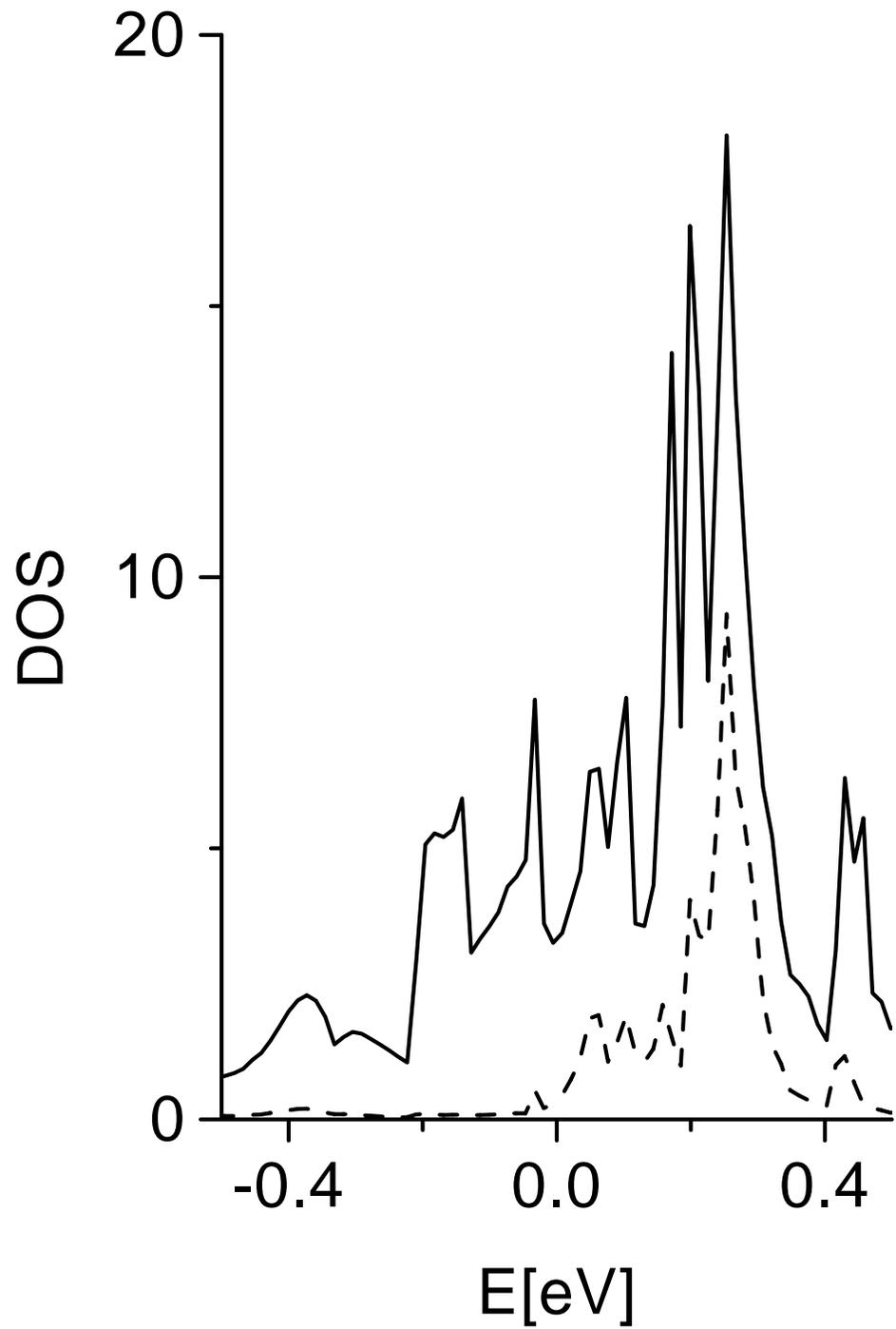
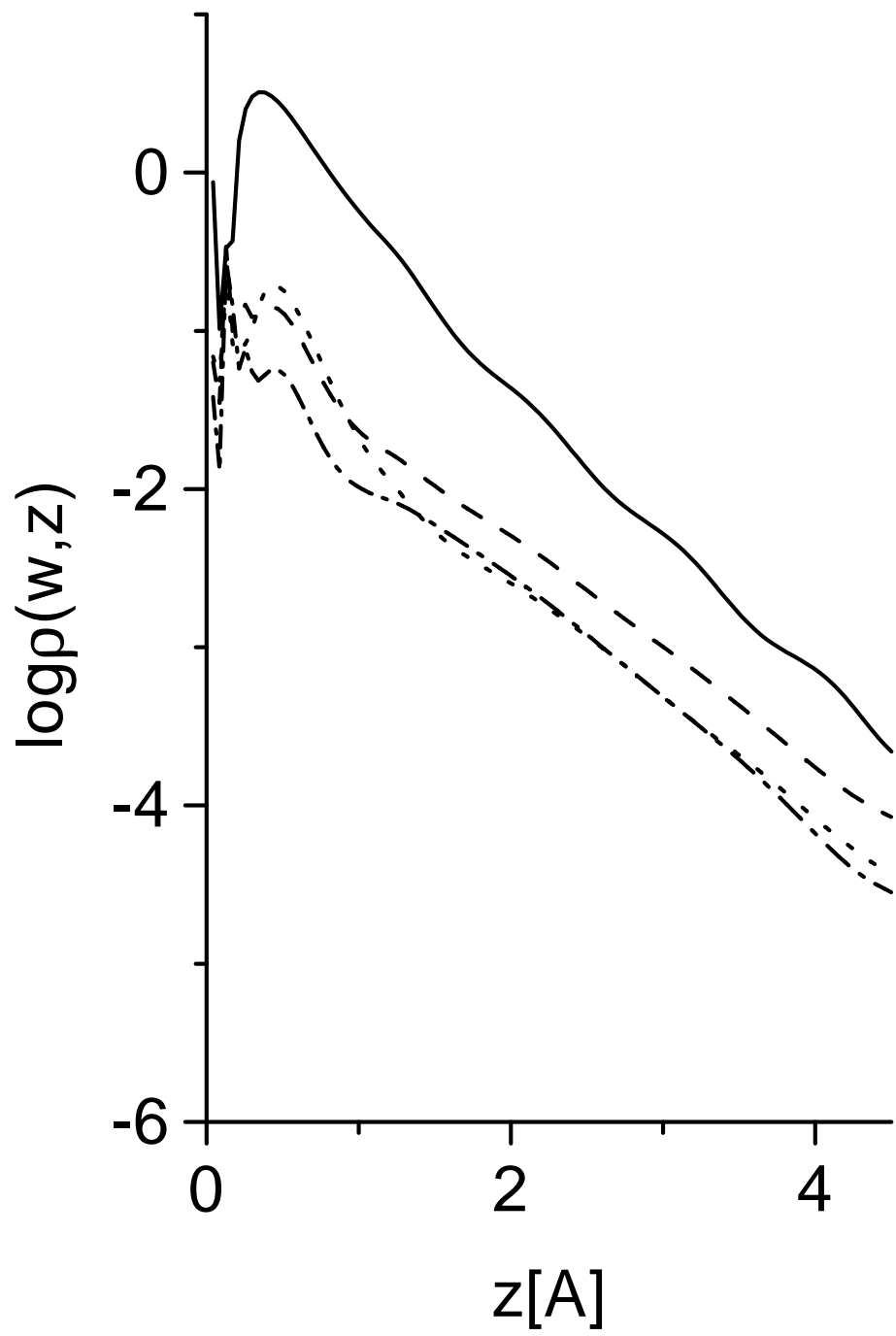
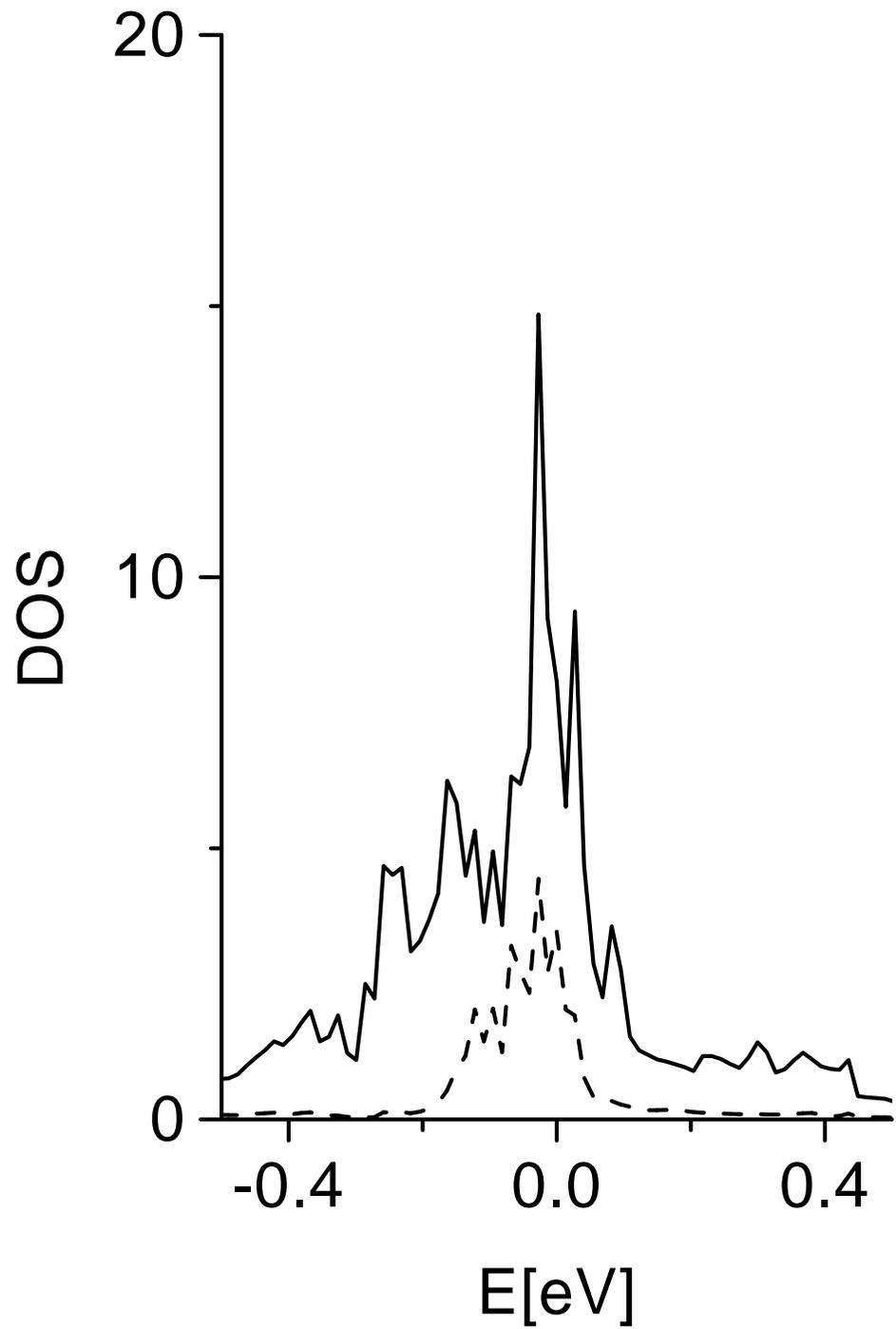
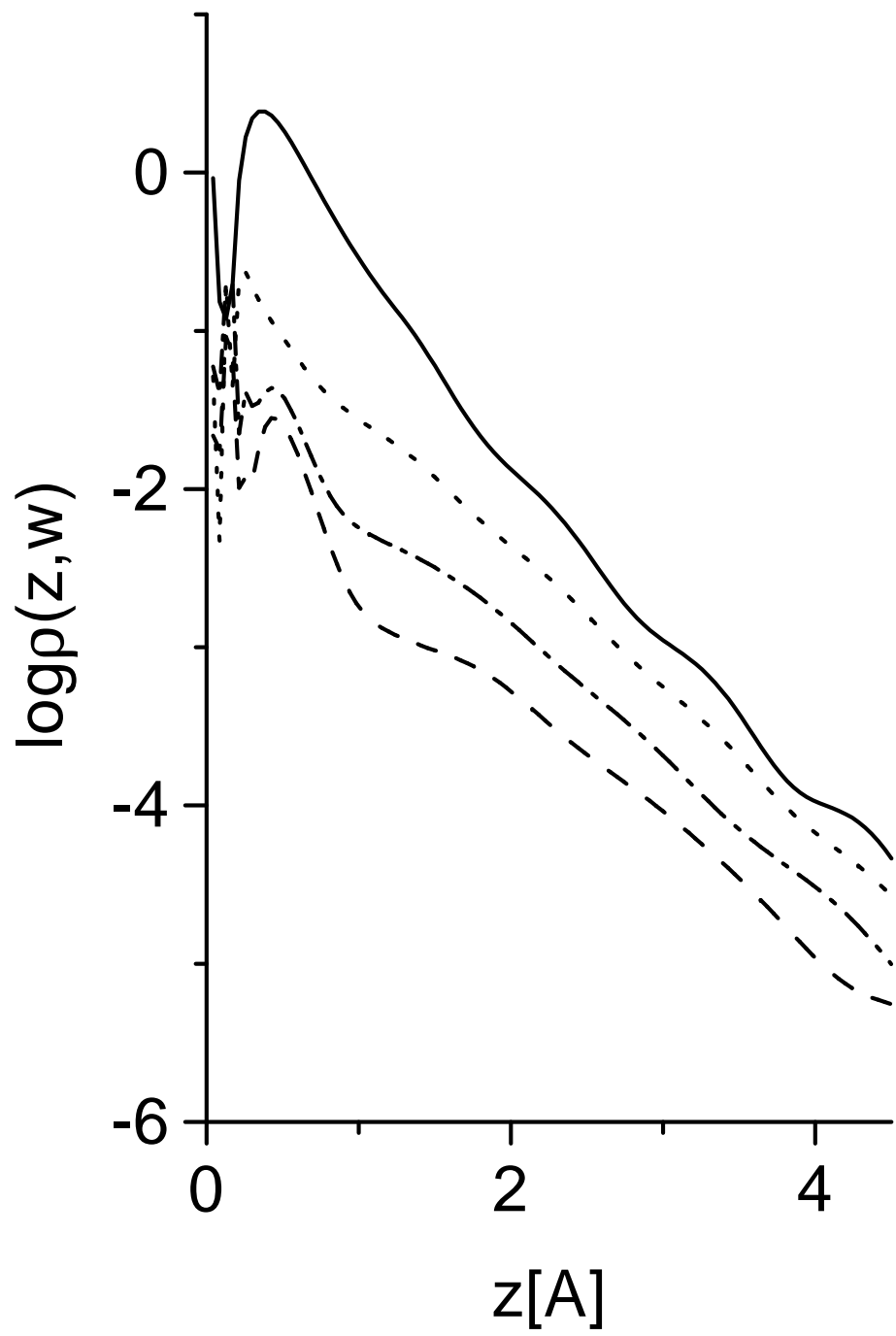
\end{document}